\documentclass{article}
\usepackage{spconf,amsmath,amssymb,amsbsy,amsxtra,graphicx}
\usepackage{array,multirow}
\usepackage{threeparttable}
\usepackage[ruled]{algorithm2e}
\usepackage{booktabs}
\usepackage{url}
\usepackage{cite}
\usepackage{psfrag}
\usepackage{dcolumn}
\usepackage{cite}
\usepackage{bm}
\usepackage{subfigure}
\usepackage{verbatim}
\usepackage{colortbl}
\usepackage{algorithmic}
\usepackage{epstopdf}

\graphicspath{{./Images/}}
\ninept
\title{A novel Adaptive weighted Kronecker Compressive Sensing}
\name{Seyed Hamid Safavi, Farah Torkamani-Azar}
\address{Cognitive Communication Research Group, \\Department of Electrical Engineering, \\ Shahid Beheshti University, Tehran, Iran}
\begin{document}
%
\maketitle
\small    
\begin{abstract}
Recently, multidimensional signal reconstruction using a low number of measurements is of great interest. Therefore, an effective sampling scheme which should acquire the most information of signal using a low number of measurements is required. In this paper, we study a novel cube-based method for sampling and reconstruction of multidimensional signals. First, inspired by the block-based compressive sensing (BCS), we divide a group of pictures (GoP) in a video sequence into cubes. By this way, we can easily store the measurement matrix and also easily can generate the sparsifying basis. The reconstruction process also can be done in parallel. Second, along with the Kronecker structure of the sampling matrix, we design a weight matrix based on the human visuality system, i.e. perceptually. We will also benefit from different weighted $\ell_1$-minimization methods for reconstruction. Furthermore, conventional methods for BCS consider an equal number of samples for all blocks. However, the sparsity order of blocks in natural images could be different and, therefore, a various number of samples could be required for their reconstruction. Motivated by this point, we will adaptively allocate the samples for each cube in a video sequence. Our aim is to show that our simple linear sampling approach can be competitive with the other state-of-the-art methods.
\end{abstract} 

\section{Introduction}\label{Introduction}
Recently, there has been a lot of interest to apply the theory of Compressive Sensing (CS)  \cite{donoho} to the high-dimensional signals such as color imaging, video, hyperspectral imaging, and multi sensor networks in a real world. In this trend, designing an efficient sampling matrix and a sparsifying basis is a challengable issue. A very simple and common approach to deal with high-dimensional data is representing them in the form of long vectors. However, this approach has some disadvantages; (a) It is not benefited from the existed correlations (or structures) between different dimensions of the data, (b) It needs high memory to store the sensing matrix and sparsifying basis, (c) The reconstruction process is computationally expensive.

To better use the structure of the multidimensional signal, Kronecker Compressive Sensing (KCS) has been proposed \cite{duarte2012kronecker}. Note that in practice, it is difficult to have global measurements of a multidimensional signal, but, partitioned measurements are available which depend only on a subset of the multidimensional signal like distributed sensor networks \cite{duarte2012kronecker}. It is also shown that the signals are more compressible in Kronecker basis than isotropic basis. Despite the valuable benefits of KCS, it has high reconstruction computational complexity for very high-dimensional data which makes it not suitable in practice. There are also some works that use tensor decomposition for designing the sparsifying basis and measurement matrix \cite{sidiropoulos2012multi, friedland2014compressive}. In \cite{sidiropoulos2012multi}, an approach called Multi-Way Compressed Sensing (MWCS) was proposed in which the CS ideas is generalized from the linear to the multi-linear case. Moreover, Generalized Tensor Compressive Sensing (GTCS) is another approach which preserves the intrinsic structure of the tensor data \cite{friedland2014compressive}. However, all of the aforementioned approaches are computationally complex compared to multidimensional approaches based on BCS.

The idea of BCS is to overcome some challenges of high-dimensional signals reconstruction \cite{fowler2012block} such as huge memory requirement to store the large-size measurement matrices and computationally expensive reconstruction process. In conventional CS methods, all frequencies of the image have the same importance. Reweighted $\ell_1$-minimization (RWL1) approach \cite{candes2008} has been proposed to enhance the sparsity by adding some weights to the signal entries. We believe that weighted compressive sensing (WCS) is a powerful method to improve the efficiency of KCS approach. The WCS approach of \cite{candes2008} utilizes an iterative algorithm to determine weights as a function of the reconstructed signal in the previous iteration. Practically the $\ell_1$-norm is replaced by log-sum penalty. In this research, motivated by the structure of the human eye, we propose to give different levels of importance to different entries of a signal in transformed domain. Since the sensitivity of human eye is not same for all frequencies, we can greatly improve the performance by assigning importance to each entries of a signal. In contrast to RWL1 approach \cite{candes2008}, since our scheme of obtaining the weights is not iterative, it could be different and much more efficient in terms of computational complexity.

To summarize, our contribution in this research is twofold: (1) First, to make the KCS approach practical to use, it is developed to cube-based one which can be processed in parallel;
(2) Second, we propose to assign some novel perceptual weights to each voxel.
We also expect the following contributions could improve the performance incredibly:
(1) We will add popular reweighted CS reconstruction methods to further gain the performance;
(2) It is well known that a priori knowledge on sparsity order of signal is not available in practice. Therefore, the upper bound on sparsity order is used. However, the necessary number of samples might be smaller than the obtained upper bound. Hence, as our future work, we can allocate a distinct number of samples for each cube \cite{Safavi2016}.

\section{Proposed Approach}

Since the RWL1 approach \cite{candes2008} is iterative and it has a higher computational complexity compared to unweighted CS, in this research, we try to answer this question: how we could design the \textit{non-iterative} and \textit{image independent} weighting matrix of CS? Motivated by the human visual system (HVS) which state that the sensitivity of human eye is not same for different frequencies, we could give an importance to each frequency in signal reconstruction. Hence, we could expect better visuality performance from the proposed scheme compared to other state-of-the-art WCS methods. One benefit that could be expected for this approach is that the weighting coefficients are image independent which makes it non-iterative and fast.
Here, we propose to take advantage from the well-known $8 \times 8$ standard luminance quantization matrix of JPEG scheme \footnote{https://www.w3.org/Graphics/JPEG/itu-t81.pdf} (which is based on the psychovisual thresholding) to design the weighting coefficients. 
Now, let us explain how we design the weighting matrix from the popular JPEG quantization matrix. We choose the inverse of the weights for each video frame in cubes as follows:

\begin{equation}
\label{proposed_weight}
\mathbf{Q'} = \mathbf{D} \mathbf{Q}^T \mathbf{D}, \qquad \mathbf{W}^{-1} = \text{Diag} \left(  \text{vec} \left(  \mathbf{Q'} \right) \right)
\end{equation}
where $\mathbf{Q}$ is the quantization matrix and $\mathbf{D}$ is an anti-diagonal matrix which is used for up-down and left-right flips. Similar to the conventional WCS methods higher weights are assigned to small values of the signal to make the signal sparser. Note that for the higher block-size, we could use interpolation for designing the weights. We develop this weight matrix for each cube in the video signal by repeating this weights along the time dimension. Let us consider the reweighted $\ell_1$-minimization problem for $j$'th cube as follows:
\begin{eqnarray}
\label{xx}
\underset{\mathbf{s}\in {{\mathbb{R}}^{N}}}{\mathop{\min }}\,\,{{\left\| {{\mathbf{W}}_j}{\mathbf{s}_j} \right\|}_{{{\ell }_{1}}}} \,\, \textmd{s.t.} \,\, \mathbf{y}_j=\bm{\theta}_j \mathbf{s}_j \quad \left( {j = 1,...,L} \right)
\end{eqnarray}
where ${\mathbf{W}}_j = \mathbf{I}_{GoP} \otimes \mathbf{W}$ and $\bm{\theta}_j$ is defined based on Kronecker product ${\bm{\theta}} = {\mathbf{\Phi}} {\mathbf{\Psi}} = {{\mathbf{\Phi }}_t} {{\mathbf{\Psi}}_t} \otimes {{\mathbf{\Phi }}_s} {{\mathbf{\Psi}}_s} = {\bm{\theta}_t} \otimes {\bm{\theta}_s}$ \cite{duarte2012kronecker}. Here ${\mathbf{\Phi}}$ denotes the sampling matrix and ${\mathbf{\Psi}}$ represents the sparsifying basis.

\section{Experiments}
We provide some simulations to validate the effectiveness of our algorithm. Different from \cite{duarte2012kronecker}, we assume Gaussian measurement matrix and DCT matrix as a sparsifying basis. First, to investigate the effectiveness of our proposed WCS scheme, we apply it to images. Fig. 1 (a) compares the PSNR of our proposed WCS algorithm with other state-of-the-art methods like iterative reweighted least square (IRLS) \footnote{we have used the MATLAB code provided by the authors and available at: http://www-m15.ma.tum.de/Allgemeines/NewPublications} and, RWL1 \cite{candes2008} method for Lenna test image. Second, we evaluate the proposed cube-based WKCS approach for video sequences. Fig. 1 (b) shows the PSNR for the proposed cube-based approach for the Foreman video sequence. It can be seen that by assigning the weights based on HVS, the performance of KCS is improved. Unlike \cite{duarte2012kronecker}, since we have assumed cube-based KCS, then assumption of dense sensing matrix (i.e. $\mathbf{\Phi}_3$) which improves the performance is reasonable. We also observed that increasing the cube size reduces the blocking artifact and improves the performance, but, the computational complexity is increased. 



\begin{figure}[t]
\begin{minipage}[b]{.5\linewidth}
  \centering
  \centerline{\includegraphics[width=4.8cm,height=4.5cm]{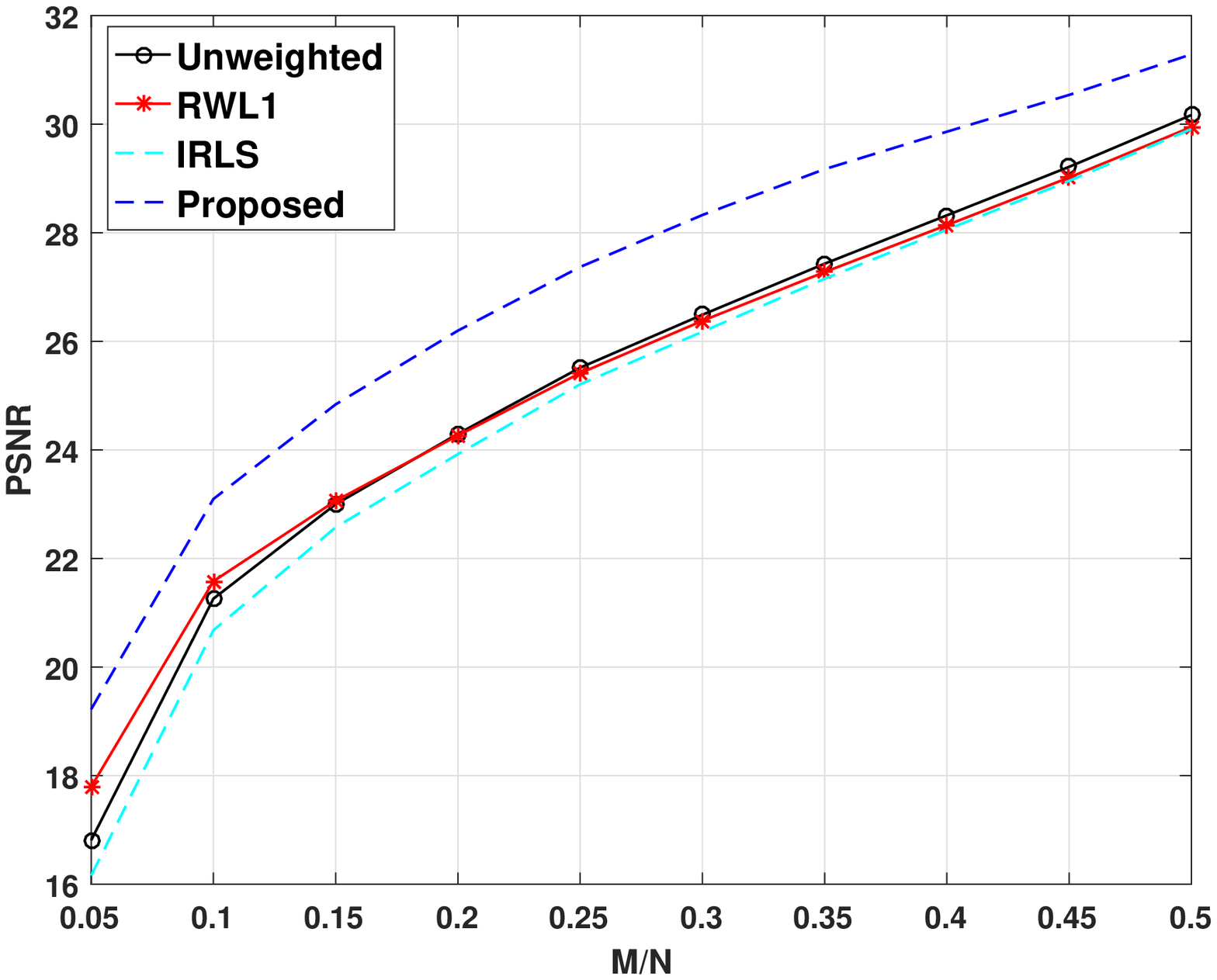}}
  \centerline{(a)}\medskip
\end{minipage}
\begin{minipage}[b]{0.5\linewidth}
  \centering
  \centerline{\includegraphics[width=4.8cm,height=4.5cm]{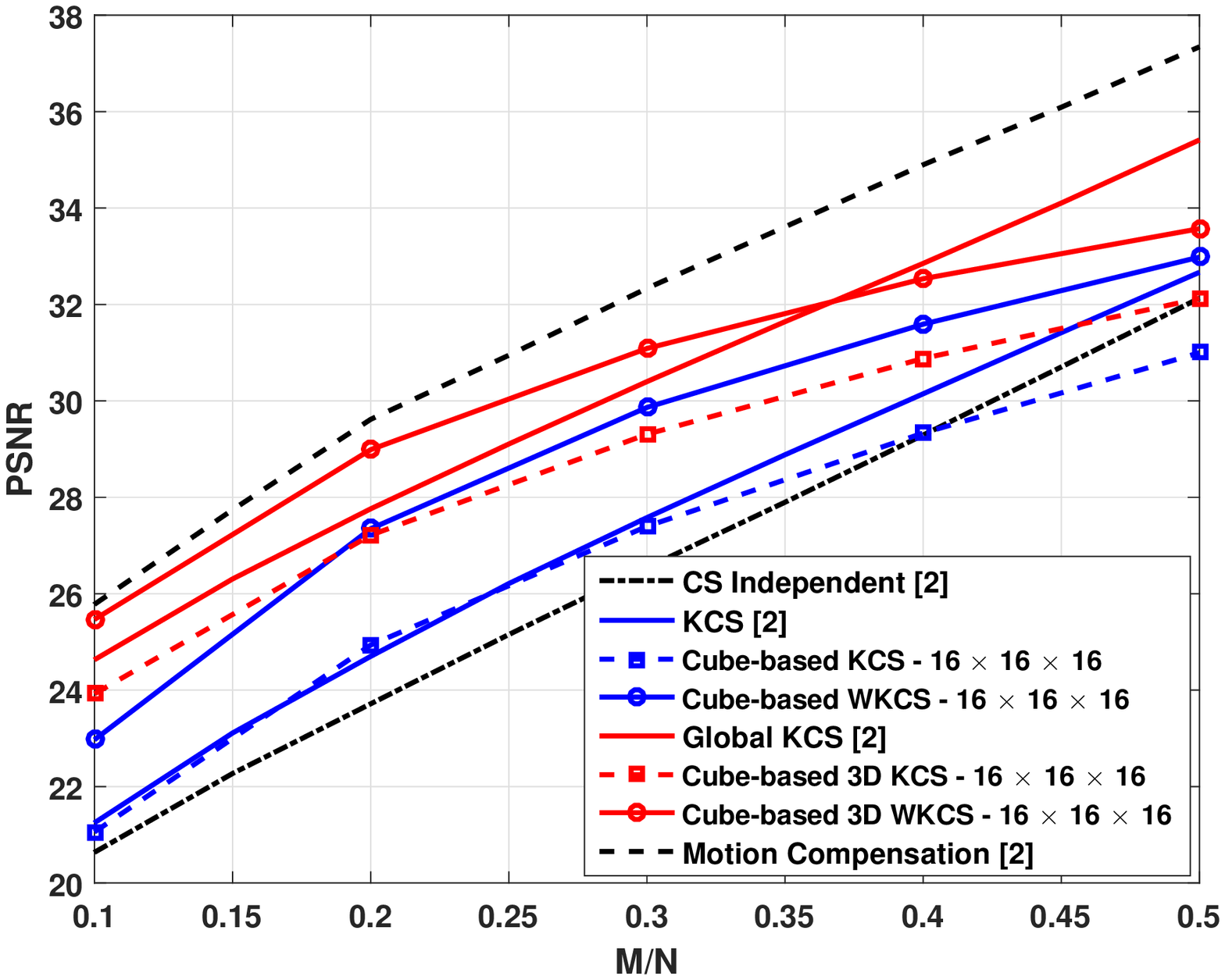}}
  \centerline{(b)}\medskip
\end{minipage}
\caption{PSNR vs normalized measurement rate: (a) Proposed WCS approach for the Lenna test image. (b) Proposed cube-based WKCS for the Foreman video sequence.}
\label{Vis}
\end{figure}

\begin{table}\small 
	\begin{tabular}{|*{1}{p{130pt}|}*{2}{c|}}
		\hline
		Methods&$\Phi$&$\Psi$\\
		\hline
		CS Independent& $\mathbf{I} \otimes \mathbf{\Phi}_s$ & $\mathbf{I} \otimes \mathbf{\Psi}_s$ \\
		\hline
		KCS, Cube-base KCS and WKCS & $\mathbf{I} \otimes \mathbf{\Phi}_s$ & $\mathbf{\Psi}_t \otimes \mathbf{\Psi}_s$\\
		\hline
		Global KCS & $\mathbf{\Phi}_t \otimes \mathbf{\Phi}_s$ & $\mathbf{\Psi}_t \otimes \mathbf{\Psi}_s$\\
		\hline
		Cube-based 3D KCS and WKCS & $\mathbf{\Phi}_3$ & $\mathbf{\Psi}_t \otimes \mathbf{\Psi}_s$\\		
		\hline
	\end{tabular}
	\centering
	\vspace{+0.1cm}
	\normalsize
	\caption{Structure of measurement and sparsifying basis matrices. }
	\label{tab:SynResult}
\end{table}
\section{Discussion and Conclusions}

In this paper, we proposed a novel cube-based weighted $\ell_{1}$ minimization algorithm based on HVS for reconstruction of compressible video signals using KCS approach. First, to make the KCS approach practical to use, it is developed to cube-based one which can be processed in parallel and also makes it applicable to any large-size videos. Second, we propose to assign some novel weights to each voxel perceptually based on the quantization matrix of JPEG. Simulation results demonstrated the superiority of our proposed WCS approach. As a future work, we propose to add popular reweighted CS reconstruction methods to further gain the performance. Tensor decomposition could also be used for processing of each cube. This could reduce the reconstruction complexity \cite{friedland2014compressive}. Adaptively allocation of the samples for each cube based on the sparsity order of each cube in video sequence could also be applied. Designing perceptual weights based on the contrast sensitivity function (CSF) for wavelet domain is also another interesting direction of this research.



\end{document}